\documentclass[a4paper,10pt]{article}
\usepackage{amsfonts}
\usepackage{amsmath}
\usepackage{amssymb}
\usepackage{graphicx}

\begin{document}

\title{Landau level transitions in doped graphene in a time dependent
magnetic field}
\author{J.S. Ardenghi$^{\dag }$\thanks{%
email:\ jsardenghi@gmail.com, fax number:\ +54-291-4595142}, P. Bechtold$%
^{\dag }$, P. Jasen$^{\dag }$, E. Gonzalez$^{\dag }$ and O. Nagel$^{\dag }$ 
\\
$^{\dag }$IFISUR, Departamento de F\'{\i}sica (UNS-CONICET)\\
Avenida Alem 1253, Bah\'{\i}a Blanca, Buenos Aires, Argentina}
\maketitle

\begin{abstract}
The aim of this work is to describe the Landau levels transitions of Bloch
electrons in doped graphene with an arbitrary time dependent magnetic field
in the long wavelength approximation. In particular, transitions from the $m$
Landau level to the $m\pm 1$ and $m\pm 2$ Landau levels are studied using
time-dependent perturbation theory. Time intervals are computed in which
transition probabilities tend to zero at low order in the coupling constant.
In particular, Landau level transitions are studied in the case of Bloch
electrons travelling in the direction of the applied magnetic force and the
results are compared with classical and revival periods of electrical
current in graphene. Finally, current probabilities are computed for the $%
n=0 $ and $n=1$ Landau levels showing expected oscillating behavior with
modified cyclotron frequency.
\end{abstract}

\section{Introduction}

Nowadays, graphene, the most important crystalline forms of carbon, is one
of the most significant topics in solid state physics due to the vast
application in nano-electronics, opto-electronics, superconductivity and
Josephson junctions (\cite{novo},\cite{intro1},\cite{intro2})\footnote{%
For a complete review of the topic see \cite{B} and \cite{BBBB}.}. This
material is a two-dimensional sheet of carbon atoms forming a honey-comb
lattice made of two interpenetrating-bonded triangular sublattices, A and B.
The linear band dispersion at the so called Dirac point is a special feature
of the graphene band structure which are dictated by the $\pi $ and $\pi
^{\prime }$ bands that form conical valleys touching at the high symmetry
points of the Brillouin zone \cite{A}. A key point is that the linear
dispersion near the symmetry points have striking similarities with those of
massless relativistic Dirac fermions or an effective Dirac-Weyl Hamiltonian 
\cite{B}. This leads to a number of fascinating phenomena such as the
half-quantized Hall effect (\cite{C},\cite{D}) and minimum quantum
conductivity in the limit of vanishing concentration of charge carriers \cite%
{novo}. Another of the effects that change their form, comparing to the
electrons described by Schr\"{o}dinger equation, are the Landau levels.
These are quantized energy levels for electrons in a magnetic field. They
still appear also for relativistic electrons, just their dependence on field
and quantization parameter is different. In a conventional non-relativistic
electron gas, Landau quantization produces equidistant energy levels, which
is due to the parabolic dispersion law of free electrons. In graphene, the
electrons have relativistic dispersion law, which strongly modifies the
Landau quantization of the energy and the position of the levels. In
particular, they are not equidistant as in conventional case and the largest
energy separation is between the zero and the first Landau level (for
multilayer graphene see \cite{AAA}). This large gap allows one to observe
the quantum Hall effect in graphene, even at room temperature \cite{E}.
Experimentally, the Landau levels in graphene have been observed by
measuring cyclotron resonances of the electrons and holes in infrared
spectroscopy experiments \cite{F} and by measuring tunneling current in
scanning tunneling spectroscopy experiments \cite{G}. The experimental study
of the $n=0$ Landau level has attracted a great deal of attention, and has
evolved rapidly, as better quality samples become available, and higher
magnetic fields and lower temperatures were studied (\cite{G1},\cite{G2},%
\cite{G3}, \cite{G3.1}), but in particular, Landau level mixing is neglected
(see \cite{Landauu}, \cite{landauu1}), and this is not as clearly justified
because the single particle Landau level gaps in graphene scale as $\sqrt{B}$%
, which is the same as the interaction strength \cite{Landau2}. In turn,
only optical transitions between Landau levels has been calculated. Since in
graphene both conduction and valence bands have the same symmetry, the
optical transition selection rule has the same form for both intraband and
interband transition, and is given by the relation $n=m\pm 1$ (see \cite{sa}
and \cite{intra}).

In the other side, it is commonly accepted that graphene has very few
lattice defects. Doped graphene shows remarkable high field-effect
mobilities, even at room temperatures (\cite{G4},\cite{G5}). Also it is
possible the appearance of negative conductivity in graphene with impurities
in magnetic fields (see \cite{mm} and \cite{BB}).

In turn, Landau level shift in graphene monolayer subjected to a quantizing
perpendicular magnetic field under the influence of short-range $\delta $%
-potential impurities has been studied using Green function method \cite{qw}%
. However, Landau level transitions in graphene with impurities placed in a
time-dependent magnetic field has not been theoretically studied, therefore
in this work we address this problem using time-dependent perturbation
theory, where the free Hamiltonian is defined on eq.(2) of \cite{BB}, and
the interacting Hamiltonian is defined in eq.(\ref{ss3}) of this work. In
particular, the case in which at $t=0$ the quantum state is in the $m$
Landau level and the possible transition to the $m\pm 1$ and $m\pm 2$ Landau
level is studied and its relation to revivals and zitterbewegung effect is
analized (see \cite{PRBB}).

The work is organized as follows:

In Section 2, the free Hamiltonian and the interacting Hamiltonian are
introduced and the time-dependent perturbation theory is applied.

In Section 3, transition probabilities from $m$ to $m\pm 1$ and $m\pm 2$
Landau levels are computed and time intervals in which probabilities goes to
zero are obtained in terms of the Hamiltonian parameters.

In Section 4, transitions from the $n=0~$Landau level to the $n=1\ $Landau
level are computed and analyzed. Current probability is obtained showing the
expected circular motion with cyclotron frequency.

In section 5, the conclusions are presented.

In Appendix A and C the conservation of probability and current density is
computed and in Appendix B some useful formulas are introduced.

\section{Graphene Hamiltonian with impurity placed in a time-dependent
magnetic field}

The Hamiltonian of graphene near the Dirac point $K$ in the long wavelength
approximation reads\footnote{%
Some of the arguments of this section are based on the work \cite{BB}, which
will be used as a starting point for the subsequent development.}%
\begin{equation}
H=v_{f}(\sigma _{x}p_{x}+\sigma _{y}p_{y})  \label{gg1}
\end{equation}%
where $v_{f}$ is the Fermi velocity that is defined as $v_{f}=\frac{\sqrt{3}%
ta}{2\hbar }$, where $t\approx 3eV$ and $a=0.246nm$ (see \cite{1}), $\sigma
_{x}$ and $\sigma _{y}$ are the Pauli matrices and $p_{x}$ and $p_{y}$ are
the momentum of the Bloch electrons in the graphene plane monolayer). This
Hamiltonian acts on the electron wavefunctions localized on sublattice $A$
and $B$ respectively. Using the tight binding approximation, is not
difficult to show that if we place an impurity near the atom of the
sublattice $A$ and $B$, the corresponding Hamiltonian can be written as (see 
\cite{BB})%
\begin{equation}
H=\left( 
\begin{array}{cccc}
0 & v_{f}(p_{x}-ip_{y}) & U & 0 \\ 
v_{f}(p_{x}+ip_{y}) & 0 & 0 & U \\ 
U & 0 & \varepsilon & 0 \\ 
0 & U & 0 & \varepsilon%
\end{array}%
\right)  \label{gg2}
\end{equation}%
where $U$ is the hybridization potential which reflects the overlap energy
between the $p_{z}$ orbital of the electron in carbon atom and the arbitrary
orbital of the impurity atom, and $\varepsilon $ is the energy of the
absorbed impurity atom with respect to the Fermi level (these values can be
evaluated experimentally (see \cite{AA})). The electron wavefunctions will
be a four component vector $\psi =\left( \rho _{A},\rho _{B},\upsilon
_{A},\upsilon _{B}\right) $, where the first two components correspond to $K$
Dirac point\ and the remaining two components correspond to the wave
function of electrons localized on the impurity.

If we place the graphene perpendicular to a time dependent magnetic field,
the momentum of the Bloch electrons will be changed as $\overrightarrow{p}%
\rightarrow \overrightarrow{p}-e\overrightarrow{A}$ where $\overrightarrow{A}
$ is the potential vector and $e$ is the electron charge. In particular, we
can choose the following gauge $\overrightarrow{A}=(-B(t)y,0,0).$ In this
case, the Hamiltonian reads%
\begin{equation}
H=\left( 
\begin{array}{cccc}
0 & v_{f}(p_{x}+eB(t)y-ip_{y}) & U & 0 \\ 
v_{f}(p_{x}+eB(t)y+ip_{y}) & 0 & 0 & U \\ 
U & 0 & \varepsilon & 0 \\ 
0 & U & 0 & \varepsilon%
\end{array}%
\right)  \label{gg3}
\end{equation}

In the following, we will assume that the magnetic field will have the
following time dependence

\begin{equation}
B(t)=B_{0}+\lambda B_{I}(t)  \label{ss1}
\end{equation}%
where $\lambda \leq 1$ is a dimensionless coupling constant and where $%
B_{I}(t)$ is an arbitrary time-dependent magnetic field. Introducing eq.(\ref%
{ss1}) on eq.(\ref{gg3}) we can separate the Hamiltonian in two parts $%
H=H_{0}+\lambda V(t)$, where

\begin{equation}
H_{0}=\left( 
\begin{array}{cccc}
0 & v_{f}(p_{x}+eB_{0}y-ip_{y}) & U & 0 \\ 
v_{f}(p_{x}+eB_{0}y+ip_{y}) & 0 & 0 & U \\ 
U & 0 & \varepsilon & 0 \\ 
0 & U & 0 & \varepsilon%
\end{array}%
\right)  \label{ss2}
\end{equation}%
and%
\begin{equation}
V(t)=\left( 
\begin{array}{cccc}
0 & v_{f}eB_{I}(t)y & 0 & 0 \\ 
v_{f}eB_{I}(t)y & 0 & 0 & 0 \\ 
0 & 0 & 0 & 0 \\ 
0 & 0 & 0 & 0%
\end{array}%
\right)  \label{ss3}
\end{equation}

The wave function can be written as $\psi =e^{-ik_{x}x}\left( \psi _{A},\psi
_{B},\xi _{A},\xi _{B}\right) $ where $k_{x}\sim 1/a$ in order to describe
the low-energy excitations, i.e. electronic excitations where the
charateristic energy may concentrate on excitations at the Fermi level and $%
\psi _{A}(\psi _{B})$ and $\xi _{A}(\xi _{B})$ are functions that depend on $%
y$. If we make the following coordinate transformation%
\begin{equation}
\overline{y}=-\hbar k_{x}+eB_{0}y  \label{gg4}
\end{equation}%
and then we make a scale transformation on $\overline{y}$ as $\overline{y}=%
\overline{\overline{y}}\,\sqrt{e\hbar B_{0}}$, then the free Hamiltonian
reads%
\begin{equation}
H_{0}=\left( 
\begin{array}{cccc}
0 & \gamma a^{\dag } & U & 0 \\ 
\gamma a & 0 & 0 & U \\ 
U & 0 & \varepsilon & 0 \\ 
0 & U & 0 & \varepsilon%
\end{array}%
\right)  \label{gg5}
\end{equation}%
where $\gamma =v_{f}\sqrt{2e\hbar B_{0}}$ and $a$ and $a^{\dag }$ are
creation and annihilation operators%
\begin{equation}
a=\frac{1}{\sqrt{2}}\left( \overline{\overline{y}}+\frac{d}{d\overline{%
\overline{y}}}\right) \text{ \ \ \ \ \ \ \ \ \ \ \ }a^{\dag }=\frac{1}{\sqrt{%
2}}\left( \overline{y}-\frac{d}{d\overline{\overline{y}}}\right)  \label{gg6}
\end{equation}%
that satisfy the commutation relations $\left[ a,a^{\dag }\right] =I$.

The eigenvalues and eigenstates for the Hamiltonian of eq.(\ref{gg5}) then
reads%
\begin{equation}
E_{n}^{(i,j)}=\frac{1}{2}\left( \varepsilon +(-1)^{i}\gamma \sqrt{n}+(-1)^{j}%
\sqrt{(\gamma \sqrt{n}+(-1)^{i+1}\varepsilon )^{2}+4U^{2}}\right)
\label{gg8}
\end{equation}%
where $i=1$, $j=1$ gives the energy levels of holes (valence band) in the
impurity atom, $i=1$, $j=2$ the energy levels of electrons (conduction band)
in the impurity atom, $i=2$, $j=1$ the energy levels of holes in the carbon
atom and $i=2$, $j=2$ the energy levels of electrons in carbon atom. The
impurities introduce a band gap $\Delta E=\sqrt{\varepsilon ^{2}+4U^{2}}$,
which implies that doped graphene becomes a semiconductor. The associated
eigenvectors reads%
\begin{equation}
\varphi _{n}^{(i,j)}(\overline{\overline{y}})=\left( 
\begin{array}{c}
\alpha _{n}^{(i,j)}\phi _{n}(\overline{\overline{y}}) \\ 
(-1)^{i}\alpha _{n}^{(i,j)}\phi _{n-1}(\overline{\overline{y}}) \\ 
(-1)^{i}\phi _{n}(\overline{\overline{y}}) \\ 
\phi _{n-1}(\overline{\overline{y}})%
\end{array}%
\right)  \label{gg9}
\end{equation}%
where $\phi _{n}$ is the eigenstate of the quantum harmonic oscillator and%
\begin{equation}
\alpha _{n}^{(i,j)}=\frac{1}{2U}\left( \gamma \sqrt{n}+(-1)^{i+1}\varepsilon
+(-1)^{i+j}\sqrt{(\gamma \sqrt{n}+(-1)^{i+1}\varepsilon )^{2}+4U^{2}}\right)
\label{gg9.1}
\end{equation}

With the knowledge of the eigenfunctions of $H_{0}$, time-dependent
perturbation theory can be applied, where the perturbation is $\lambda V(t)$%
. In particular it will be assumed that from $t=-\infty $ to $t=0$ the
magnetic field is $B(t)=B_{0}$ and from $t=0$ to $t=+\infty $, $%
B(t)=B_{0}+\lambda B_{I}(t)$. In this sense, the basis of eigenvectors of $%
H_{0}$ can be written as the direct sum of four subspaces, each for energy
band. Each subspace is expand by a infinite basis defined by the harmonic
oscillator eigenfunctions. In this work we will study the possible
transitions within an energy band, which in algebraic terms implies that we
will expand the wave function of the full Hamiltonian in terms of the basis
of one subspace of the direct sum.

\subsection{Time dependent perturbation theory}

As we said in the previous section we will study the possible transitions of
the different Landau levels in the conduction band of electrons in carbon
atoms, then $i=2$ and $j=2$.\footnote{%
In the following we will disregard these two labels. Is not difficult to
show that there are not transitions between different energy bands with the
interaction Hamiltonian $V(t)$.}

The wave function of the Bloch electrons in graphene near the impurity in
the long wave-length approximation can be written as%
\begin{equation}
\psi (x,\overline{\overline{y}},t)=\sum\limits_{n=0}^{+\infty
}c_{n}(t)e^{-ik_{x}x}\varphi _{n}(\overline{\overline{y}})e^{-\frac{i}{\hbar 
}E_{n}t}  \label{ss5}
\end{equation}%
where the $c_{n}(t)$ coefficients represent the probability amplitude of the
perturbed quantum system in the $n$-state (see Appendix A). The matrix
equation for the coefficients reads%
\begin{equation}
i\hbar \frac{dc_{m}}{dt}=\lambda \sum\limits_{n=0}^{+\infty }\left( \int
\varphi _{m}^{\ast }(\overline{\overline{y}})V(t)\varphi _{n}(\overline{%
\overline{y}})d\overline{\overline{y}}dx\right) e^{-i\omega
_{(n,m)}t}c_{n}(t)  \label{ss6}
\end{equation}%
where%
\begin{equation}
\omega _{(n,m)}=\frac{1}{\hbar }(E_{n}-E_{m})=\frac{1}{2\hbar }\left( \gamma
\left( \sqrt{n}-\sqrt{m}\right) +\sqrt{(\gamma \sqrt{n}-\varepsilon
)^{2}+4U^{2}}-\sqrt{(\gamma \sqrt{m}-\varepsilon )^{2}+4U^{2}}\right)
\end{equation}%
are the spectral frequencies. Applying $V(t)$ on $\varphi _{n}(\overline{%
\overline{y}})$ and computing the scalar product with $\varphi _{m}^{\ast }(%
\overline{\overline{y}})$ we obtain%
\begin{gather}
\int \varphi _{m}^{\ast }(\overline{\overline{y}})V(t)\varphi _{n}(\overline{%
\overline{y}})d\overline{\overline{y}}dx=-\frac{\alpha _{n}\alpha _{m}\eta
(t)}{\sqrt{(\alpha _{n}^{2}+1)(\alpha _{m}^{2}+1)}}[\beta \left( \delta
_{m,n-1}+\delta _{m-1,n}\right)  \label{ss7} \\
+\frac{\xi \eta (t)}{\sqrt{2}}\left( \sqrt{n-1}\delta _{m,n-2}+\sqrt{n}%
\delta _{mn}+\sqrt{n}\delta _{m-1,n-1}+\sqrt{n+1}\delta _{m-1,n+1}\right) ] 
\notag
\end{gather}%
where $\beta =\,v_{f}\sqrt{e\hbar B_{0}}=\sqrt{2}\gamma $, $\xi =v_{f}\hbar
k_{x}$, $\eta (t)=\frac{B_{I}(t)}{B_{0}}$ and where it has been used that $%
v_{f}eB_{I}(t)y=\eta (t)(\beta \overline{\overline{y}}+\xi )$ and $\overline{%
\overline{y}}=\frac{1}{\sqrt{2}}(a+a^{\dag })$. Introducing last equation on
eq.(\ref{ss6})\ we obtain a coupled system of differential equations for the 
$c_{m}~$coefficients%
\begin{gather}
i\hbar \frac{dc_{m}}{dt}=-\lambda (\frac{\alpha _{m}^{2}\eta (t)}{\alpha
_{m}^{2}+1}\xi \sqrt{2m}c_{m}(t)+\frac{\alpha _{m+1}\alpha _{m}\eta (t)}{%
\sqrt{(\alpha _{m+1}^{2}+1)(\alpha _{m}^{2}+1)}}\beta e^{-i\omega
_{(m+1,m)}t}c_{m+1}(t)  \label{ss8} \\
+\frac{\alpha _{m-1}\alpha _{m}\eta (t)}{\sqrt{(\alpha _{m-1}^{2}+1)(\alpha
_{m}^{2}+1)}}\beta e^{-i\omega _{(m-1,m)}t}c_{m-1}(t)+\frac{\alpha
_{m+2}\alpha _{m}\eta (t)}{\sqrt{(\alpha _{m+2}^{2}+1)(\alpha _{m}^{2}+1)}}%
\xi \sqrt{\frac{m+1}{2}}e^{-i\omega _{(m+2,m)}t}c_{m+2}(t)  \notag \\
+\frac{\alpha _{m-2}\alpha _{m}\eta (t)}{\sqrt{(\alpha _{m-2}^{2}+1)(\alpha
_{m}^{2}+1)}}\xi \sqrt{\frac{m-1}{2}}e^{-i\omega _{(m-2,m)}t}c_{m-2}(t)) 
\notag
\end{gather}%
Last equation indicates that the contribution to the $m$-state comes from
the $m-2$, $m-1$, $m$, $m+1$ and $m+2$ states.

The perturbation approximation is introduced by expanding the coefficients
in powers of $\lambda $ (see eq.(12.48) of \cite{Ballentine}, page 350),
then at order $\lambda ^{0}$ and order $\lambda ^{k}$ we obtain the
following relations among the coefficients%
\begin{equation}
\frac{dc_{m}^{(0)}}{dt}=0  \label{ss81.}
\end{equation}%
and%
\begin{gather}
i\hbar \frac{dc_{m}^{(k)}}{dt}=-\frac{\alpha _{m}^{2}}{\alpha _{m}^{2}+1}\xi 
\sqrt{2m}\eta (t)c_{m}^{(k-1)}(t)-\frac{\alpha _{m+1}\alpha _{m}}{\sqrt{%
(\alpha _{m+1}^{2}+1)(\alpha _{m}^{2}+1)}}\beta e^{-i\omega _{(m+1,m)}t}\eta
(t)c_{m+1}^{(k-1)}(t)  \label{ss9} \\
-\frac{\alpha _{m-1}\alpha _{m}}{\sqrt{(\alpha _{m-1}^{2}+1)(\alpha
_{m}^{2}+1)}}\beta e^{-i\omega _{(m-1,m)}t}\eta (t)c_{m-1}^{(k-1)}(t)  \notag
\\
-\frac{\alpha _{m+2}\alpha _{m}}{\sqrt{(\alpha _{m+2}^{2}+1)(\alpha
_{m}^{2}+1)}}\xi \sqrt{\frac{m+1}{2}}e^{-i\omega _{(m+2,m)}t}\eta
(t)c_{m+2}^{(k-1)}(t)  \notag \\
-\frac{\alpha _{m-2}\alpha _{m}}{\sqrt{(\alpha _{m-2}^{2}+1)(\alpha
_{m}^{2}+1)}}\xi \sqrt{\frac{m-1}{2}}e^{-i\omega _{(m-2,m)}t}\eta
(t)c_{m-2}^{(k-1)}(t)  \notag
\end{gather}%
where the superscript in $c_{m}^{(k)}$ indicates the order of $\lambda $ and
the subscript indicates the Landau level. In the following section we will
study the possible transitions from the $m$ state to the $m\pm 1$ and $m\pm
2 $ states under different conditions up to second order in the perturbation
expansion.

\section{Transition probabilities from $m$ to $m\pm 1$ and $m\pm 2$ Landau
levels}

Using eq.(\ref{ss9})\ and assuming that at $t=0$ the quantum state is in the 
$m$ state, that is $c_{m}^{(0)}=1$ and $c_{k}^{(0)}=0$ for $k\neq m$, we
obtain for $c_{m}^{(1)}(t)$, $c_{m-1}^{(1)}(t)$, $c_{m+1}^{(1)}(t)$, $%
c_{m-2}^{(1)}(t)$ and $c_{m+2}^{(1)}(t)$ the following equations%
\begin{equation}
c_{m}^{(1)}=-\frac{\alpha _{m}^{2}}{i\hbar (\alpha _{m}^{2}+1)}\xi \sqrt{2m}%
f_{(0,0)}(t)  \label{tt1}
\end{equation}%
\begin{equation}
c_{m+1}^{(1)}=-\frac{\alpha _{m}\alpha _{m+1}}{i\hbar \sqrt{(\alpha
_{m}^{2}+1)(\alpha _{m+1}^{2}+1)}}\beta f_{(m,m+1)}(t)  \label{tt2}
\end{equation}%
\begin{equation}
c_{m-1}^{(1)}=-\frac{\alpha _{m}\alpha _{m-1}}{i\hbar \sqrt{(\alpha
_{m}^{2}+1)(\alpha _{m-1}^{2}+1)}}\beta f_{(m,m-1)}(t)  \label{tt3}
\end{equation}%
\begin{equation}
c_{m+2}^{(1)}=-\frac{\alpha _{m}\alpha _{m+2}}{i\hbar \sqrt{(\alpha
_{m}^{2}+1)(\alpha _{m+2}^{2}+1)}}\xi \sqrt{\frac{m+1}{2}}f_{(m,m+2)}(t)
\label{tt4}
\end{equation}%
\begin{equation}
c_{m-2}^{(1)}=-\frac{\alpha _{m}\alpha _{m-2}}{i\hbar \sqrt{(\alpha
_{m}^{2}+1)(\alpha _{m-2}^{2}+1)}}\xi \sqrt{\frac{m-1}{2}}f_{(m,m-2)}(t)
\label{tt5}
\end{equation}%
where%
\begin{equation}
f_{(n,m)}(t)=\int_{0}^{t}\eta (t^{\prime })e^{-i\omega _{(n,m)}t^{\prime
}}dt^{\prime }  \label{tt6}
\end{equation}%
and $c_{k}^{(1)}=0$ for $k<m-2$ and $k>m+2$. From eq.(\ref{tt4}) and eq.(\ref%
{tt5}), it can be noted that the $c_{m\pm 2}^{(2)}$ coefficients depend on $%
\xi $, which is a function of the wave vector $k_{x}$. Because we will study
the transition probabilities up to order $\lambda ^{2}$, we must compute the 
$c_{m}^{(2)}$ coefficient which reads%
\begin{gather}
c_{m}^{(2)}=-\frac{\alpha _{m}^{2}}{\hbar ^{2}(\alpha _{m}^{2}+1)}[\xi
^{2}\left( \frac{2m\alpha _{m}^{2}}{(\alpha _{m}^{2}+1)}F_{(0,0)}^{(0,0)}(t)+%
\frac{\alpha _{m+2}^{2}(m+1)}{2(\alpha _{m+2}^{2}+1)}%
F_{(m,m+2)}^{(m+2,m)}(t)+\frac{\alpha _{m-2}^{2}(m-1)}{2(\alpha _{m-2}^{2}+1)%
}F_{(m,m-2)}^{(m-2,m)}(t)\right)  \label{tt7} \\
+\beta ^{2}\left( \frac{\alpha _{m+1}^{2}}{(\alpha _{m+1}^{2}+1)}%
F_{(m,m+1)}^{(m+1,m)}(t)+\frac{\alpha _{m-1}^{2}}{(\alpha _{m-1}^{2}+1)}%
F_{(m,m-1)}^{(m-1,m)}(t)\right) ]  \notag
\end{gather}%
where%
\begin{equation}
F_{(n,m)}^{(n^{\prime },m^{\prime })}(t)=\int_{0}^{t}e^{-i\omega
_{(n^{\prime },m^{\prime })}t^{\prime }}\eta (t^{\prime
})f_{(n,m)}(t^{\prime })dt^{\prime }  \label{tt8}
\end{equation}%
Using the Taylor expansion of eq.(\ref{tt6}) and eq.(\ref{tt8}) in Appendix
B, the probabilities up to order $\lambda ^{2}$ and order $t^{2}$ reads%
\footnote{%
The contributions at first and second order in time-dependent perturbation
theory are valid for small values of $t$.}%
\begin{gather}
P_{m}(t)=1-\frac{\lambda ^{2}\alpha _{m}^{2}\eta _{0}^{2}t^{2}}{\hbar
^{2}(\alpha _{m}^{2}+1)}\times  \label{tt9} \\
\left[ \xi ^{2}\left( \frac{2m\alpha _{m}^{2}}{\alpha _{m}^{2}+1}+\frac{%
\alpha _{m+2}^{2}(m+1)}{2(\alpha _{m+2}^{2}+1)}+\frac{\alpha _{m-2}^{2}(m-1)%
}{2(\alpha _{m-2}^{2}+1)}\right) +\beta ^{2}\left( \frac{\alpha _{m+1}^{2}}{%
(\alpha _{m+1}^{2}+1)}+\frac{\alpha _{m-1}^{2}}{(\alpha _{m-1}^{2}+1)}%
\right) \right] +O(\lambda ^{3},t^{3})  \notag
\end{gather}%
for $P_{m\pm 1}(t)$ we obtain%
\begin{equation}
P_{m+1}(t)=\lambda ^{2}\frac{\alpha _{m}^{2}\alpha _{m+1}^{2}\beta ^{2}}{%
\hbar ^{2}(\alpha _{m}^{2}+1)(\alpha _{m+1}^{2}+1)}\eta
_{0}^{2}t^{2}+O(\lambda ^{3},t^{3})  \label{tt10}
\end{equation}%
\begin{equation}
P_{m-1}(t)=\lambda ^{2}\frac{\alpha _{m}^{2}\alpha _{m-1}^{2}\beta ^{2}}{%
\hbar ^{2}(\alpha _{m}^{2}+1)(\alpha _{m-1}^{2}+1)}\eta
_{0}^{2}t^{2}+O(\lambda ^{3},t^{3})  \label{tt11}
\end{equation}%
and for $P_{m\pm 2}(t)$ we obtain%
\begin{equation}
P_{m+2}(t)=\lambda ^{2}\frac{\alpha _{m}^{2}\alpha _{m+2}^{2}\xi ^{2}(m+1)}{%
2\hbar ^{2}(\alpha _{m}^{2}+1)(\alpha _{m+2}^{2}+1)}\eta
_{0}^{2}t^{2}+O(\lambda ^{3},t^{3})  \label{tt12}
\end{equation}%
\begin{equation}
P_{m-2}(t)=\lambda ^{2}\frac{\alpha _{m}^{2}\alpha _{m-2}^{2}\xi ^{2}(m-1)}{%
2\hbar ^{2}(\alpha _{m}^{2}+1)(\alpha _{m-2}^{2}+1)}\eta
_{0}^{2}t^{2}+O(\lambda ^{3},t^{3})  \label{tt13}
\end{equation}%
where $\eta _{0}=B_{I}(0)/B_{0}$ is a dimensionless constant. From last
equation is not difficult to see that $P_{m}(t)$ contains the coefficients
that appears in $P_{m+1}(t)$, $P_{m-1}(t)$, $P_{m+2}(t)$ and $P_{m-2}(t)$,
that is

\begin{equation}
P_{m}(t)=1-\lambda ^{2}\frac{2m\alpha _{m}^{4}\eta _{0}^{2}t^{2}\xi ^{2}}{%
\hbar ^{2}(\alpha _{m}^{2}+1)^{2}}%
-P_{m+2}(t)-P_{m-2}(t)-P_{m+1}(t)-P_{m-1}(t)  \label{tt13.1}
\end{equation}%
The second term of r.h.s. of last equation, which is the contribution at
first order in $\lambda $ of the probability amplitude of the $m$ Landau
level, prevents that Bloch electrons behave as a five-level system in the
following time interval%
\begin{equation}
\Delta t=\frac{\hbar \sqrt{\alpha _{m}^{2}+1}}{\lambda \alpha _{m}\eta _{0}%
\sqrt{\xi ^{2}\left( \frac{2m\alpha _{m}^{2}}{\alpha _{m}^{2}+1}+\frac{%
\alpha _{m+2}^{2}(m+1)}{2(\alpha _{m+2}^{2}+1)}+\frac{\alpha _{m-2}^{2}(m-1)%
}{2(\alpha _{m-2}^{2}+1)}\right) +\beta ^{2}\left( \frac{\alpha _{m+1}^{2}}{%
(\alpha _{m+1}^{2}+1)}+\frac{\alpha _{m}^{2}}{(\alpha _{m-1}^{2}+1)}\right) }%
}  \label{tt14}
\end{equation}%
which is defined as the time at which $P_{m}(t)$ is zero. The limits $%
m\rightarrow 0$ reads%
\begin{equation}
\underset{m\rightarrow 0}{\lim }\Delta t=\frac{2\hbar \sqrt{4u^{2}+\epsilon
(\epsilon +\sqrt{4u^{2}+\epsilon ^{2}})}}{(\epsilon +\sqrt{4u^{2}+\epsilon
^{2}})\eta \beta \sqrt{\frac{(\epsilon +\sqrt{4u^{2}+\epsilon ^{2}})^{2}}{%
4u^{2}+\epsilon (\epsilon +\sqrt{4u^{2}+\epsilon ^{2}})}+\frac{2(\sqrt{z}%
\beta +\sqrt{2}\epsilon +\sqrt{8u^{2}+z\beta ^{2}+2\sqrt{z}\beta \epsilon
+2\epsilon ^{2}})^{2}}{8u^{2}+(\sqrt{z}\beta +\sqrt{2}\epsilon +\sqrt{%
8u^{2}+z\beta ^{2}+2\sqrt{z}\beta \epsilon +2\epsilon ^{2}})^{2}}}}
\label{tt14.01}
\end{equation}

\begin{figure}[tbp]
\centering
\includegraphics[width=100mm,height=65mm]{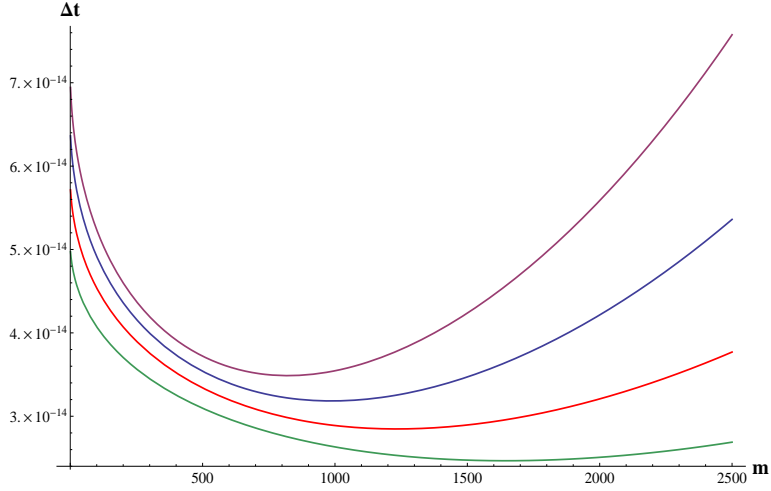}
\caption{$\Delta t$ as a function of $m$ for $z=3$(violet), $z=4$ (blue), $%
z=5$ (red) and $z=6$ (green).}
\label{figura1.2}
\end{figure}

In particular, using typical values, for example, $U=-1.43eV$, $\varepsilon
=-3eV$, $\xi =9/2eV$, $\gamma =\frac{\beta }{\sqrt{2}}=0.043\sqrt{\frac{z}{2}%
}eV$, where $z$ is a dimensioneless parameter that quantifies the magnetic
field strength $B_{0}$ (we use $z=5$) and $B_{I}(0)=5T$, so $\eta _{0}=1$,
the limit of $\Delta t$ when $m\rightarrow 0$ is about $\Delta t\sim
1.55\times 10^{-15}s$, which is in the order of the optical spectrum for
electromagnetic waves. This value can be compared with the classical period
and the revival time of electron current in graphene under a constant
magnetic field (see \cite{PRBB}), which do not depends on the presence or
ausence of impurities. Both values $T_{CL}=\frac{4\pi }{v_{f}}\sqrt{\frac{%
m\hbar }{2eB_{0}}}$ and $T_{R}=\frac{16\pi m^{3/2}}{v_{f}}\sqrt{\frac{\hbar 
}{2eB_{0}}}$ are of the order of $10^{-12}s$ for $B_{0}=10T$ and where $m$
is the Landau level index. The main difference between the values found in
this section and the approach of \cite{PRBB} is that in the latter a
superposition of two wave packets, one from the conduction band and the
other from the valence band, are used. In turn, the zitterbewegung effect,
which is related with the interference between positive and negative
frequency components of wave packets superposition of both positive and
negative energies shows up with a period $T_{ZB}=\pi \sqrt{\frac{\hbar }{%
m2eB_{0}}}$. In \cite{PRBB} values of $T_{ZB}$ are of the order of $%
10^{-15}s $ which coincide with the values computed in eq.(\ref{tt14.01})).
The similar values for $\Delta t$ and $T_{ZB}$ implies that we must consider
the contribution of the valence band to the wave function superposition of
eq.(\ref{ss5}). In that case, $V(t)$ will not introduce coupling effects
between the conduction and valence band transitions amplitudes and a similar
Landau level transitions will appear in the valence band. In the next
section we will show how the $T_{ZB}$ can be related with $\Delta t$ in a
more direct way. In figure 1, $\Delta t$ is plotted against $m$ for various
values of $z$. The figure shows relevant initial Landau levels where $\Delta
t$ reaches a minimum, which is the shortest time scale that characterizes
the spread of the $m$ Landau level energy to the $m\pm 1$~and $m\pm 2$
Landau level energies as a function of $m$. Using that $\Delta E\sim
\left\vert
E_{m_{0}+1}+E_{m_{0}-1}+E_{m_{0}+2}+E_{m_{0}-2}-E_{m_{0}}\right\vert $,
where $m_{0}$ is the Landau level at which $\Delta t$ reaches the minimum,
the uncertainty principle saturate at the time interval $\Delta T(m_{0})\sim 
\frac{\hbar }{2\Delta E(m_{0})}$ which is smaller than $\Delta t$; using the
values above $\Delta T(m_{0})/\Delta t(m_{0})\sim 10^{-3}$. This implies
that we cannot obtain a coherent state in the time interval in which the $m$%
, $m\pm 1$ and $m\pm 2$ Landau levels get mixed.

The ratio between the considered Landau levels reads%
\begin{equation}
\frac{P_{m-1}(t)}{P_{m+1}(t)}=\frac{\alpha _{m-1}^{2}(\alpha _{m+1}^{2}+1)}{%
\alpha _{m+1}^{2}(\alpha _{m-1}^{2}+1)}  \label{tt14.1}
\end{equation}%
and%
\begin{equation}
\frac{P_{m-2}(t)}{P_{m+2}(t)}=\frac{\alpha _{m-2}^{2}(m-1)(\alpha
_{m+2}^{2}+1)}{\alpha _{m+2}^{2}(m+1)(\alpha _{m-2}^{2}+1)}  \label{tt14.2}
\end{equation}%
In figure 1, both ratio between probabilities are plotted using the values
used above.

\begin{figure}[tbp]
\centering
\includegraphics[width=100mm,height=65mm]{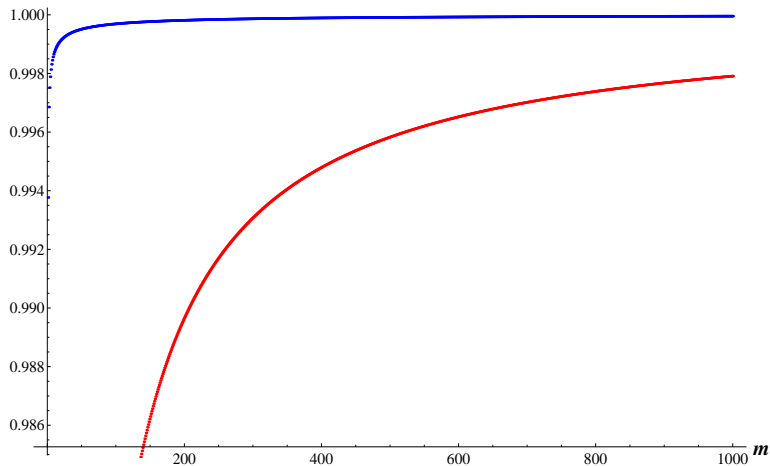}
\caption{$\frac{P_{m-1}(t)}{P_{m+1}(t)}$ (blue) and $\frac{P_{m-2}(t)}{%
P_{m+2}(t)}$ (red) as function of Landau level index $m$.}
\label{figura1.2}
\end{figure}
Both ratio probabilities have the same limit $\underset{m\rightarrow \infty }%
{\lim }\frac{P_{m-1}(t)}{P_{m+1}(t)}=\underset{m\rightarrow \infty }{\lim }%
\frac{P_{m-2}(t)}{P_{m+2}(t)}=1$ and from figure 1, $P_{m-1}(t)<P_{m+1}(t)$
and $P_{m-2}(t)<P_{m+2}(t)$, which implies that the initial probability
flows from the $m$ state to the $m+1$ and $m+2$ states faster than to the $%
m-1$ and $m-2$ states.

\subsection{Bloch electrons travelling in the $y$ direction ($k_{x}=0$)
\qquad}

In this case, $\xi =0$, then the $P_{m\pm 2}=0$ at order $\lambda ^{2}$
because $c_{m\pm 2}^{(1)}(t)=0$ (see eq.(\ref{tt4}) and eq.(\ref{tt5})), $%
P_{m\pm 1}$ remains the same and $P_{m}(t)$ reads%
\begin{equation}
P_{m}(t)=1-\frac{\lambda ^{2}\alpha _{m}^{2}\eta _{0}^{2}t^{2}\beta ^{2}}{%
\hbar ^{2}(\alpha _{m}^{2}+1)}\left( \frac{\alpha _{m+1}^{2}}{(\alpha
_{m+1}^{2}+1)}+\frac{\alpha _{m-1}^{2}}{(\alpha _{m-1}^{2}+1)}\right)
\label{k1}
\end{equation}%
From last equation and eq.(\ref{tt10}) and eq.(\ref{tt11}) we obtain the
following relation between probabilities 
\begin{equation}
P_{m}(t)+P_{m+1}(t)+P_{m-1}(t)=1  \label{k2}
\end{equation}%
in the time interval%
\begin{equation}
\Delta t=\frac{\hbar }{\lambda \alpha _{m}\beta \eta _{0}}\sqrt{\frac{%
(\alpha _{m}^{2}+1)(\alpha _{m+1}^{2}+1)(\alpha _{m-1}^{2}+1)}{\alpha
_{m+1}^{2}(\alpha _{m-1}^{2}+1)+\alpha _{m-1}^{2}(\alpha _{m+1}^{2}+1)}}
\label{k3}
\end{equation}%
In the case $k_{x}=0$, Bloch electrons behave as a three-level system in the
time interval defined in last equation, where the initial probability flows
from the $m$ state to the $m\pm 1$ states. The limit for $m\rightarrow
\infty $ of $\Delta t$ reads%
\begin{equation}
\underset{m\rightarrow \infty }{\lim }\Delta t=\frac{\hbar }{\sqrt{2}\beta
\eta _{0}}  \label{k3.1}
\end{equation}%
Using typical values, this limit is about $\Delta t\sim 10^{-13}\frac{B_{0}}{%
B_{I}(0)}s$ which is again in the order of optical spectrum for $\frac{B_{0}%
}{B_{I}(0)}\sim 1$. This time can be larger in those cases in which $B_{0}$
is bigger than $B_{I}(0)$. In this case, $\Delta t$ do not reaches a minimum
for same $m$ value. The ratio between $\Delta t$ of eq.(\ref{k3.1}) and $%
T_{CL}$ and $T_{R}$ is $T_{CL}/\Delta t=4\pi \sqrt{m}\eta _{0}$ and $%
T_{R}/T_{CL}=16\pi m^{3/2}\eta _{0}$, which coincide when $m=(4\pi \eta
_{0})^{-2}$ and $m=(16\pi \eta _{0})^{-2/3}$. In figure 2 the probability $%
P_{m}(t)$ and $P_{m+1}$ is shown at order $\lambda ^{2}$ for several values
of $m$. The probability $P_{m-1}$ can be deduced from eq.(\ref{k2}).

\begin{figure}[tbp]
\centering
\includegraphics[width=110mm,height=75mm]{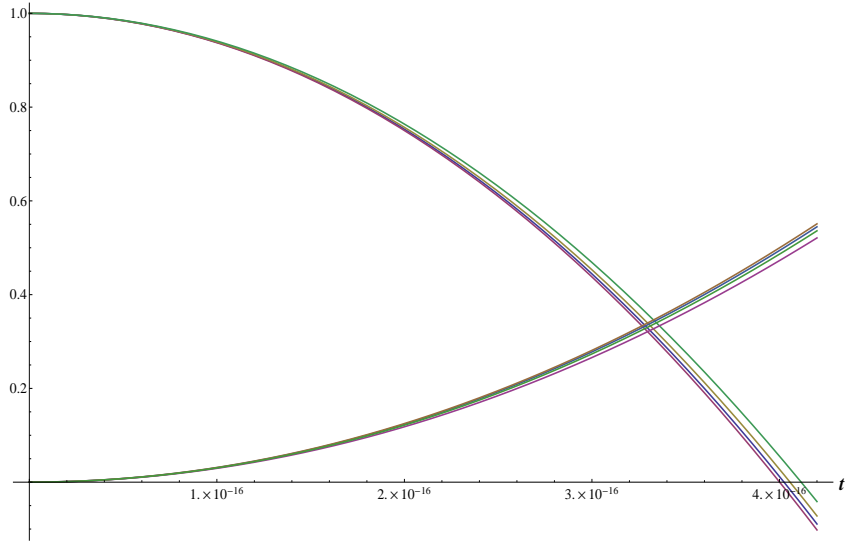}
\caption{$P_{m}(t)$ and $P_{m+1}$ probabilities as a function of time (for
larger $m$, $P_{m}(t)$ goes to zero more faster).}
\label{figura1.2}
\end{figure}
A final remark about the probability is the $m\rightarrow \infty $ limit in
eq.(\ref{ss8})\ in the case $\xi =0$%
\begin{equation}
\underset{m\rightarrow \infty }{\lim }\frac{\alpha _{m+1}\alpha _{m}\beta }{%
i\hbar \sqrt{(\alpha _{m+1}^{2}+1(\alpha _{m}^{2}+1))}}\eta (t)e^{-i\omega
_{(m+1,m)}t}=\frac{\beta }{i\hbar }\eta (t)  \label{li7}
\end{equation}%
and%
\begin{equation}
\underset{m\rightarrow \infty }{\lim }\frac{\alpha _{m-1}\alpha _{m}\beta }{%
i\hbar \sqrt{(\alpha _{m-1}^{2}+1(\alpha _{m}^{2}+1))}}\eta (t)e^{-i\omega
_{(m-1,m)}t}=\frac{\beta }{i\hbar }\eta (t)  \label{li8}
\end{equation}%
In turn if we consider 
\begin{equation}
\underset{m\rightarrow \infty }{\lim }c_{m+1}\sim \underset{m\rightarrow
\infty }{\lim }c_{m-1}\sim c_{\infty }  \label{li9}
\end{equation}%
for $m\rightarrow \infty $, eq.(\ref{ss8}) reads%
\begin{equation}
\frac{dc_{\infty }}{dt}=-\frac{2i\beta }{\hbar }\eta (t)c_{\infty }
\label{li10}
\end{equation}%
which has a trivial solution%
\begin{equation}
c_{\infty }(t)=c_{\infty }(t_{0})\exp \left( -\frac{2i\beta }{\hbar B_{0}}%
\int_{t_{0}}^{t}B_{I}(t^{\prime })dt^{\prime }\right)  \label{li10.1}
\end{equation}%
A particular case can be given when 
\begin{equation}
\int_{t_{0}}^{t}B_{I}(t^{\prime })dt^{\prime }=\mathrm{Re}(g(t))-i\mathrm{Im}%
(g(t))  \label{li10.2}
\end{equation}%
Then%
\begin{equation}
\left\vert c_{\infty }(t)\right\vert ^{2}=\left\vert c_{\infty
}(t_{0})\right\vert ^{2}e^{-\frac{4\beta }{\hbar }\mathrm{Im}(g(t))}
\label{li10.3}
\end{equation}%
which implies that a contribution to the probability is given by the
imaginary part of the integral of the time-dependent magnetic field. For
example, in an oscillating magnetic field $B_{I}(t)=B_{I}e^{-i\omega t}$,
the imaginary part of the integral of the magnetic field reads

\begin{equation}
\mathrm{Im}(g(t))=\frac{B_{I}}{\omega }(-1+\cos \omega t)  \label{li10.4}
\end{equation}%
then the probability $\left\vert c_{\infty }(t)\right\vert ^{2}$ is
oscillatory%
\begin{equation}
\left\vert c_{\infty }(t)\right\vert ^{2}=\left\vert c_{\infty
}(t_{0})\right\vert ^{2}e^{-\frac{4\beta }{\hbar }\frac{B_{I}}{\omega }%
(-1+\cos \omega t)}  \label{li10.5}
\end{equation}%
with a frequency identical to the magnetic field $B_{I}(t)$.

\section{Transition from $n=0$ to $n=1$ Landau level}

Of particular interest is the case in which the initial state is in the $n=0$
Landau level, that is, $c_{0}^{(0)}=1$ and $k_{x}=0$. The unique
probabilities that contribute at order $\lambda ^{2}$ comes from the $n=0$
and $n=1$ Landau level%
\begin{equation}
P_{0}(t)=1-\frac{\lambda ^{2}\alpha _{0}^{2}\alpha _{1}^{2}\beta ^{2}\eta
_{0}^{2}}{\hbar ^{2}(\alpha _{0}^{2}+1)(\alpha _{1}^{2}+1)}t^{2}+O(\lambda
^{3},t^{3})  \label{lll1}
\end{equation}%
and%
\begin{equation}
P_{1}(t)=\lambda ^{2}\frac{\alpha _{0}^{2}\alpha _{1}^{2}\beta ^{2}\eta
_{0}^{2}}{\hbar ^{2}(\alpha _{0}^{2}+1)(\alpha _{1}^{2}+1)}t^{2}+O(\lambda
^{3},t^{3})  \label{lll2}
\end{equation}%
In the following time interval%
\begin{equation}
\Delta t=\frac{\hbar }{\lambda \alpha _{0}\eta _{0}\beta \alpha _{1}}\sqrt{%
(\alpha _{0}^{2}+1)(\alpha _{1}^{2}+1)}  \label{lll3}
\end{equation}

In figure 3 we can see the probabilities $P_{0}(t)$ and $P_{1}(t)$ as a
function of time using the values introduced in previous section and in
Figure 4, $\Delta t$ is plotted as a function of $B_{0}$ for different
values of $B_{I}(0)$. In the time interval defined in last equation the
quantum system behaves as a two-level system. This effect could be related
to the echo effect \cite{echo}, which can be obtained with an oscillating
electric field in the $n=1$ and $n=2$ Landau level. In this case, the Landau
levels involved decay for longer times due to the coupled system of
differential equations of eq.(\ref{ss9}). Is of major interest to study
different initial conditions with different time-dependent magnetic fields
for larger orders in $\lambda $ to obtain more detailed functions for the
probability amplitudes and the conditions that allow possible cycles between
the lowest Landau levels. As we said in Section 3, if we consider the
valence band in the wave function superposition of eq.(\ref{ss5}), which
means that we have to consider $i=2$ and $j=1$ in (\ref{gg9.1}) and the
initial state is $c_{0}^{(0)}=1$, then transitions to the $n=1$ Landau
levels in the conduction and valence band will be obtained at order $\lambda
^{2}$. The time interval at which the probability flow from the initial
quantum state to these Landau levels reads%
\begin{equation}
\Delta t=\left\vert \frac{\hbar }{\lambda \alpha _{0}\eta _{0}\beta }\sqrt{%
\frac{(\alpha _{0}^{2}+1)(\alpha _{1}^{2}+1)((\frac{\gamma }{U}-\alpha
_{1})^{2}+1)}{\alpha _{1}^{2}((\frac{\gamma }{U}-\alpha _{1})^{2}+1)+(\frac{%
\gamma }{U}-\alpha _{1})^{2}(\alpha _{1}^{2}+1)}}\right\vert  \label{lll4}
\end{equation}%
which is of the order of $\Delta t\sim 10^{-14}s$, which is in concordance
with the zitterbewegung period $T_{ZB}$. Then, the zitterbewegung effect can
be understood as a natural phenomena due to the symmetry in the possible
transitions from a $m$ Landau level to the $m\pm 1$ and $m\pm 2$ Landau
levels. In the case in which we choose as initial state a superposition of
conduction and valence states, the effect will be predominant and the
correlation between the positive and negative states will be related to the
probability conservation.

\begin{figure}[tbp]
\centering
\includegraphics[width=110mm,height=75mm]{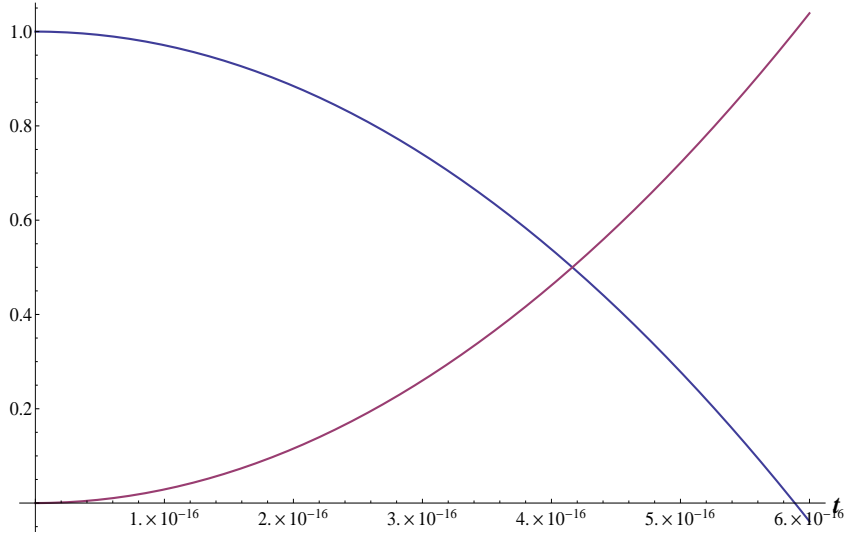}
\caption{$P_{0}(t)$ (blue) and $P_{1}(t)$ (violet) as a function of time.}
\label{figura1.2}
\end{figure}

\begin{figure}[tbp]
\centering
\includegraphics[width=100mm,height=65mm]{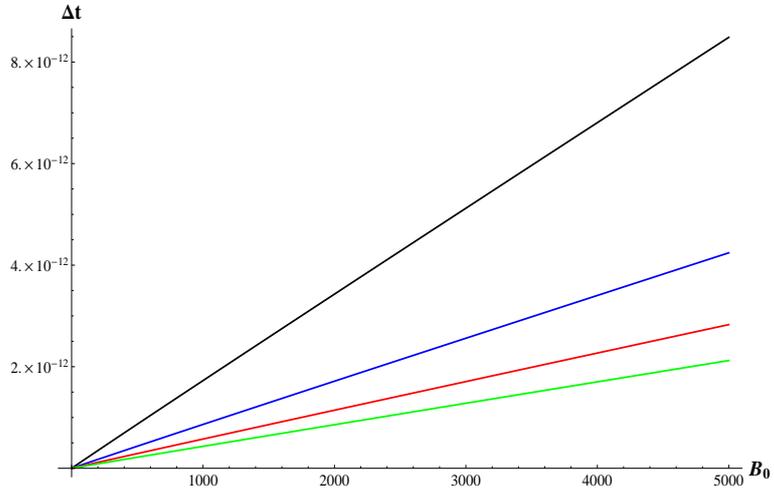}
\caption{$\Delta t$ as a function of as a function of $B_{0}$ for different
values of $B_{I}(0)$ (blue $\protect\eta =10$, black $\protect\eta =20$,
green $\protect\eta =30$, yellow $\protect\eta =40$ and red $\protect\eta %
=50 $).}
\label{figura1.2}
\end{figure}

In the other side, the probability flux can be computed using Appendix C in
the particular case in which $k_{x}=0$. The $n=0$ contribution to the
current in the $x$ and $y$ direction reads 
\begin{equation}
J_{x}^{(0)}=\lambda \frac{\alpha _{0}^{2}\alpha _{1}^{2}\beta \eta _{0}t}{%
2\hbar (\alpha _{1}^{2}+1)(\alpha _{0}^{2}+1)}\left( 1-\lambda ^{2}\frac{%
\alpha _{0}^{2}\alpha _{1}^{2}\beta ^{2}}{2\hbar ^{2}(\alpha
_{0}^{2}+1)(\alpha _{1}^{2}+1)}\eta _{0}^{2}t^{2}\right) \sin (\omega
_{(1,0)}t)  \label{li12.1}
\end{equation}

\begin{figure}[tbp]
\centering
\includegraphics[width=100mm,height=65mm]{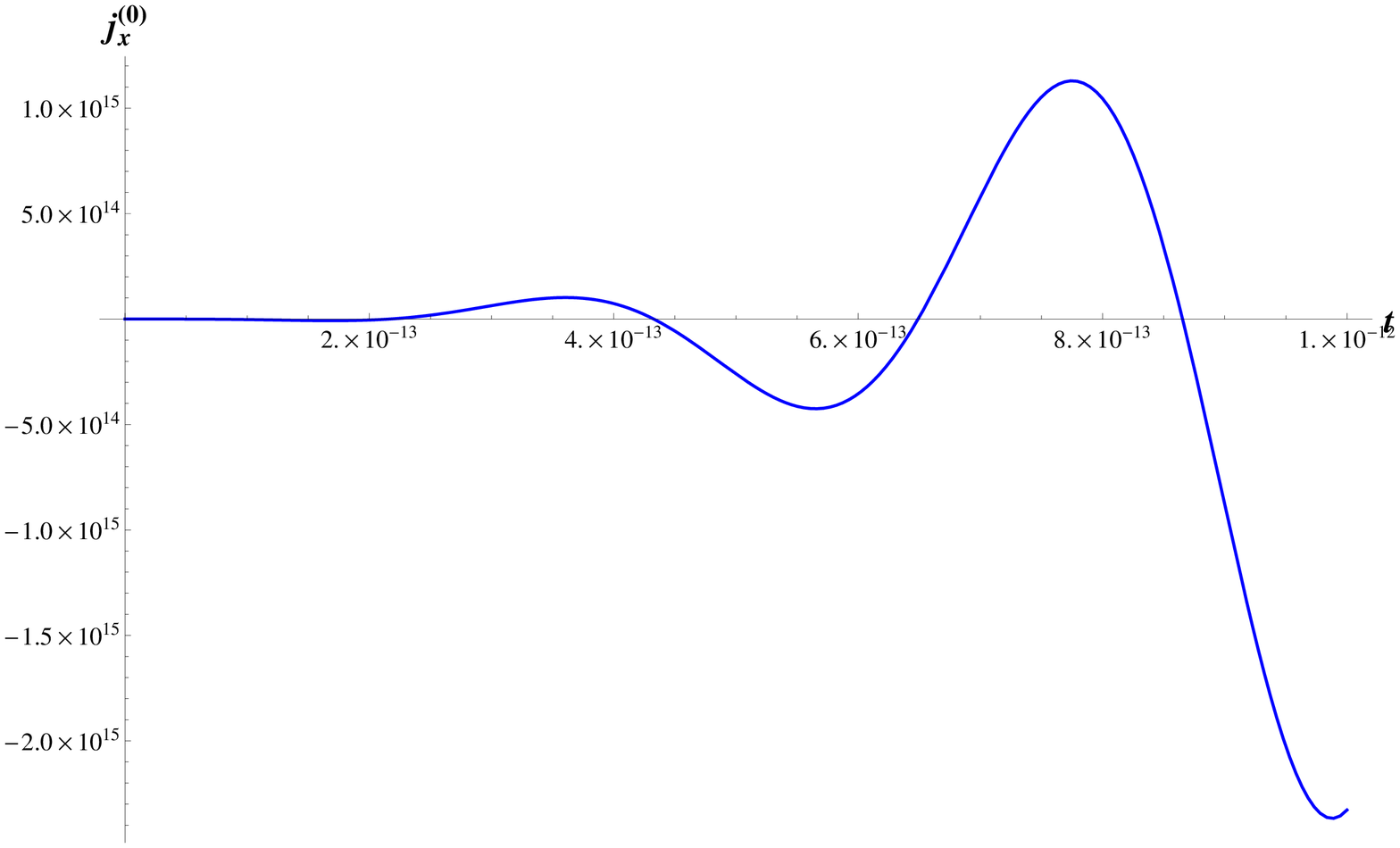}
\caption{$n=0$ contribution of the $x$ direction of the probability flux as
a function of $t$.}
\label{figure3}
\end{figure}

\begin{figure}[tbp]
\centering
\includegraphics[width=100mm,height=65mm]{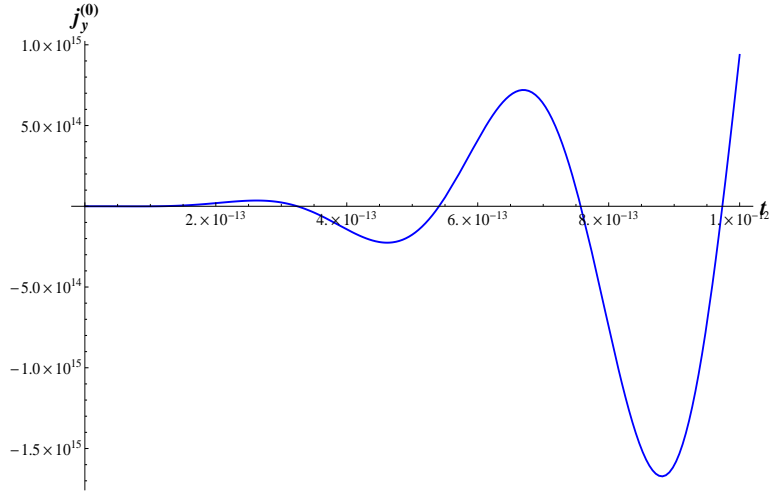}
\caption{$n=0$ contribution of the $y$ direction of the probability flux as
a function of $t$.}
\label{figure3}
\end{figure}

and for the $y$ direction%
\begin{equation}
J_{y}^{(0)}=\lambda \frac{\alpha _{0}^{2}\alpha _{1}^{2}\beta \eta _{0}t}{%
2\hbar (\alpha _{1}^{2}+1)(\alpha _{0}^{2}+1)}(1-\lambda ^{2}\frac{\alpha
_{0}^{2}\alpha _{1}^{2}\beta ^{2}}{2\hbar ^{2}(\alpha _{0}^{2}+1)(\alpha
_{1}^{2}+1)}\eta _{0}^{2}t^{2})\cos (\omega _{(1,0)}t)  \label{li14}
\end{equation}

Last results (or eq.(\ref{ppp8}) and eq.(\ref{ppp9})) can be compared with
the electronic current found in \cite{PRBB} (eq.(9)). The main difference
lies in the time dependent factor of eq.(\ref{ppp8}) and eq.(\ref{ppp9}). In
turn, the oscillating behavior depends on the spectral frequencies in both
cases, which implies that the electronic current will not depend on the
superposition of states of Bloch electrons.

From eq.(\ref{li12.1}) and eq.(\ref{li14}), the current flows in circular
motion, as it is expected, with a modified cyclotron frequency $\omega
_{(1,0)}$.\footnote{%
In the case $\epsilon =0$ and $U=0$ we obtain the cyclotron frequency $%
\omega =v_{f}\sqrt{2eB}$.} If an additional constant electric field is
applied along the doped graphene monolayer sheet, the result will be
equivalent to monolayer doped graphene subjected to a reduced magnetic field 
$\widetilde{B}(t)=B(t)\sqrt{1-\theta ^{2}}$, where $\theta =E/v_{f}B(t)$
(see \cite{libro}, page 245). In turn, the current of eq.(\ref{ppp6}) of
Appendix C will not be the net current circulating in graphene monolayer
with impurities, because a time-dependent magnetic field will induce a
time-dependent electric field that will points in the circumferential
direction%
\begin{equation}
\overrightarrow{E}=-\frac{1}{2}r\frac{dB}{dt}\widehat{e}_{\theta }
\label{li15}
\end{equation}%
This electric field will produce, from a classical viewpoint, a \ linear
current density in the $\widehat{e}_{\theta }$ 
\begin{equation}
\overrightarrow{J}=-e\sqrt{\frac{2e\pi }{m}\frac{dB}{dt}}\widehat{e}_{\theta
}  \label{li16}
\end{equation}%
pointing in the opposite direction to the current generated by the Lorentz
force. A quantum mechanical treatment is necessary to obtain the net current
probability in doped graphene under a time-dependent magnetic field. It will
be source of future works to study the current flux and probability
transitions for an oscillating magnetic field at several orders in $\lambda $%
. Finally, another line of research will be the possible interacting
Hamiltonian that allows possible transitions between different energy bands
in the long wavelength approximation.\footnote{%
A recent work is \cite{WW}, where interband transitions in bilayer graphene
in a perpendicular electric field has been studied.}

\section{Conclusions}

In this paper we have shown how to compute the Landau levels transitions of
doped graphene in a time dependent magnetic field. Starting at $t=0$ from an
arbitrary $m$ Landau level, we show the general behavior of the $m\pm 1$ and 
$m\pm 2$ Landau levels at order $\lambda ^{2}$. We study the time interval
at which the $m$ Landau level probability amplitude decrease to zero and its
relation with the parameters of the Hamiltonian, showing minimum values of $%
m $ at which the $m$ Landau level vanishes. In turn, the transition from the 
$m $ to the $m\pm 1$ Landau level in the case in which Bloch electrons
travel in the $y$ direction was analized. A three level system can be
obtained in the time interval at which the $m$ Landau level vanishes with
the difference that time interval do not reaches a minimum for some $m$
value. Finally, the $n=0$ and $n=1$ Landau level transition is studied in
the $k_{x}=0$ regime showing that for a small time interval the electrons
behave as a two-level system. The relation to the zitterbewegung effect and
revivals are analized, showing similar time intervals for both effects and
the probability flux due to the perturbation potential. Current
probabilities are computed for low orders in $\lambda $ showing an
oscillating behavior in time and where in particular negative values are
found.

\section{Acknowledgment}

This paper was partially supported by grants of CONICET (Argentina National
Research Council) and Universidad Nacional del Sur (UNS) and by ANPCyT
through PICT 1770, and PIP-CONICET Nos. 114-200901-00272 and
114-200901-00068 research grants, as well as by SGCyT-UNS., E.A.G. and
P.V.J. are members of CONICET. P.B. and J. S.A. are fellow researchers at
this institution.

The five authors are extremely grateful to the reviewer, whose relevant
observations have greatly improved the final version of this paper.

\appendix

\section{The probability amplitude}

The norm of the wave function $e^{-ik_{x}x}\varphi _{n}(\overline{\overline{y%
}})$ reads%
\begin{equation}
\int \left\vert e^{-ik_{x}x}\varphi _{n}(\overline{\overline{y}})\right\vert
^{2}dxd\overline{\overline{y}}=4L_{x}(\alpha _{n}^{2}+1)  \label{aa1}
\end{equation}%
where $2L_{x}$ is the total length in the $x$-direction of the graphene
layer. From last equation, the normalization factor reads%
\begin{equation}
N_{n}=\frac{1}{2\sqrt{L_{x}(\alpha _{n}^{2}+1)}}  \label{aa2}
\end{equation}%
Using the normalized wave function $e^{-ik_{x}x}\varphi _{n}$, the norm of
the quantum state of eq.(\ref{ss5}) reads%
\begin{equation}
\int \left\vert \psi (x,\overline{\overline{y}},t)\right\vert ^{2}dxd%
\overline{\overline{y}}=\sum\limits_{n=0}^{+\infty }\left\vert
c_{n}(t)\right\vert ^{2}=1  \label{ap2}
\end{equation}%
Then the probability amplitude for each Landau level reads%
\begin{equation}
P_{n}(t)=\left\vert c_{n}(t)\right\vert ^{2}  \label{ap4}
\end{equation}%
which will be used in Section 2.

\section{Taylor expansion}

To obtain the Taylor expansion of $f_{(n,m)}(t)$ and $F_{(n,m)}^{(n^{\prime
},m^{\prime })}(t)$ (eq.(\ref{tt6}) and eq.(\ref{tt8})) around $t=0$ we can
proceed as follows: the function $\eta (t)$ is known because depends on the
applied time-dependent magnetic field $B_{I}(t)$.\footnote{%
We can assume that it is an analytical function.} We can compute the Taylor
expansion of this function around $t=0$ 
\begin{equation}
\eta (t)=\sum\limits_{j=0}^{+\infty }\frac{\eta _{j}}{j!}t^{j}  \label{av1}
\end{equation}%
where%
\begin{equation}
\eta _{j}=\frac{d^{j}\eta (t)}{dt^{j}}(t=0)  \label{av2}
\end{equation}%
In principle we can assum that this Taylor expansion converges to $\eta (t)$
with a radius of convergence $R$, where $R\ $reads%
\begin{equation}
\frac{1}{R}=\underset{n\rightarrow \infty }{\lim }\left\vert \frac{\eta
_{j+1}}{(j+1)\eta _{j}}\right\vert  \label{av2.1}
\end{equation}%
In the other side, we can compute the Taylor expansion of $e^{-i\omega
_{(n,m)}t}$ around $t=0$%
\begin{equation}
e^{-i\omega _{(n,m)}t}=\sum\limits_{k=0}^{+\infty }\frac{(-i\omega
_{(n,m)})^{k}t^{k}}{k!}  \label{av3}
\end{equation}%
with an infinite radius of convergence. The Cauchy product of the two series
reads%
\begin{equation}
\left( \sum\limits_{j=0}^{+\infty }\frac{\eta _{j}}{j!}t^{j}\right) \left(
\sum\limits_{k=0}^{+\infty }\frac{(-i\omega _{(n,m)})^{k}t^{k}}{k!}\right)
=\sum\limits_{n=0}^{+\infty }t^{n}\sum\limits_{k=0}^{n}\frac{\eta _{k}}{k!}%
\frac{(-i\omega _{(n,m)})^{n-k}}{(n-k)!}  \label{av4}
\end{equation}%
This product converges in radius of convergence $R$.\footnote{%
If at least one of the series is absolutely convergent, then the product
must be convergent.} Finally, integrating the last result we obtain the
Taylor expansion of $f_{(n,m)}(t)$ around $t=0$ with radius of convergence $%
R $ which reads 
\begin{equation}
f(t)=\sum\limits_{n=0}^{+\infty }\frac{t^{n+1}}{n+1}\sum\limits_{k=0}^{n}%
\frac{\eta _{k}}{k!}\frac{(-i\omega _{(n,m)})^{n-k}}{(n-k)!}  \label{av5}
\end{equation}%
Because we want to compute the behavior of the $c_{m}^{(k)}$ coefficients
near $t=0$, which is the time at which the time-dependent magnetic field is
turn on, we can compute the few first terms of eq.(\ref{av5})%
\begin{equation}
f(t)=\eta _{0}t+\frac{t^{2}}{2}\left( \eta _{1}-i\omega _{(n,m)}\eta
_{0}\right) +\frac{t^{3}}{3}\left( \frac{\eta _{2}}{2}-i\omega _{(n,m)}\eta
_{1}-\eta _{0}\frac{\omega _{(n,m)}^{2}}{2}\right) +O(t^{4})  \label{av6}
\end{equation}%
where%
\begin{equation}
\eta _{0}=\eta (0)=\frac{B_{I}(0)}{B_{0}}  \label{av6.1}
\end{equation}%
\begin{equation}
\eta _{1}=\frac{d\eta }{dt}(t=0)=\frac{1}{B_{0}}\frac{dB_{I}}{dt}(t=0)
\label{av7}
\end{equation}%
and%
\begin{equation}
\eta _{2}=\frac{d^{2}\eta }{dt^{2}}(t=0)=\frac{1}{B_{0}}\frac{d^{2}B_{I}}{%
dt^{2}}(t=0)  \label{av8}
\end{equation}%
In a similar way, we can proceed with the Taylor expansion of eq.(\ref{tt8})
by using the expansion of eq.(\ref{av1}), eq.(\ref{av3}) and the result of
eq.(\ref{av5}). The product of the three expansions reads%
\begin{equation}
\eta (t)e^{-i\omega _{(n^{\prime },m^{\prime
})}t}f_{(n,m)}(t)=\sum\limits_{n=0}^{+\infty
}t^{n+1}\sum\limits_{k=0}^{n}\sum\limits_{l=0}^{k}\frac{\eta _{l}}{l!}\frac{%
(-i\omega _{(n^{\prime },m^{\prime })})^{k-l}}{\left( n-k+1\right) (k-l)!}%
\left( \sum\limits_{s=0}^{n-k}\frac{\eta _{s}}{s!}\frac{(-i\omega
_{(n,m)})^{n-k-s}}{(n-k-s)!}\right)  \label{av9}
\end{equation}%
then the integral reads%
\begin{eqnarray}
F_{(n,m)}^{(n^{\prime },m^{\prime })}(t) &=&\int_{0}^{t}e^{-i\omega
_{(n^{\prime },m^{\prime })}t^{\prime }}\eta (t^{\prime
})f_{(n,m)}(t^{\prime })dt^{\prime }=\sum\limits_{n=0}^{+\infty }\frac{%
t^{n+2}}{n+2}\times  \label{av10} \\
&&\sum\limits_{k=0}^{n}\sum\limits_{l=0}^{k}\frac{\eta _{l}}{l!}\frac{%
(-i\omega _{(n^{\prime },m^{\prime })})^{k-l}}{\left( n-k+1\right) (k-l)!}%
\left( \sum\limits_{s=0}^{n-k}\frac{\eta _{s}}{s!}\frac{(-i\omega
_{(n,m)})^{n-k-s}}{(n-k-s)!}\right)  \notag
\end{eqnarray}%
This last result is the Taylor expansion of the function $%
F_{(n,m)}^{(n^{\prime },m^{\prime })}(t)$ with convergence radius $R$. The
first terms of this Taylor expansion read

\begin{equation}
F_{(n,m)}^{(n^{\prime },m^{\prime })}(t)=\frac{\eta _{0}}{2}t^{2}+\frac{\eta
_{0}}{3}t^{3}\left( \frac{3}{2}\eta _{1}-i\eta _{0}\left( \omega
_{(n^{\prime },m^{\prime })}+\frac{\omega _{(n,m)}}{2}\right) \right)
+O(t^{4})  \label{av11}
\end{equation}%
Eq.(\ref{av6}) and eq.(\ref{av11}) will be used in Section 3.

\section{Probability flux}

The Hamiltonian of eq.(\ref{ss2}) with an electromagnetic field can be
written in a compact form as%
\begin{equation}
H=\left[ 
\begin{array}{cc}
v_{f}\overrightarrow{\sigma }\cdot (-i\hbar \overrightarrow{\nabla }-e%
\overrightarrow{A}) & UI \\ 
UI & \epsilon I%
\end{array}%
\right] \left[ 
\begin{array}{c}
\psi \\ 
\xi%
\end{array}%
\right] =i\hbar \frac{\partial }{\partial t}\left[ 
\begin{array}{c}
\psi \\ 
\xi%
\end{array}%
\right]  \label{ppp1}
\end{equation}%
where $\psi $ is a two component wave function of the electrons in carbon
atom and $\xi $ is a two component wave function of the electrons in the
impurity atom. Last equation can be separated in two coupled differential
equations for $\psi $ and $\xi $%
\begin{equation}
v_{f}\overrightarrow{\sigma }\cdot (-i\hbar \overrightarrow{\nabla }-e%
\overrightarrow{A})\psi +U\xi =i\hbar \frac{\partial \psi }{\partial t}
\label{ppp2}
\end{equation}%
and%
\begin{equation}
U\psi +\epsilon \xi =i\hbar \frac{\partial \xi }{\partial t}  \label{ppp3}
\end{equation}%
The probability density reads%
\begin{equation}
\rho =\psi ^{\dag }\psi +\xi ^{\dag }\xi  \label{ppp4}
\end{equation}%
by taking the time derivate we obtain the following continuity equation%
\begin{equation}
\frac{\partial \rho }{\partial t}+\overrightarrow{\nabla }\left( v_{f}\psi
^{\dag }\overrightarrow{\sigma }\psi \right) =0  \label{ppp5}
\end{equation}%
Then, the current probability reads%
\begin{equation}
\overrightarrow{j}=v_{f}\psi ^{\dag }\overrightarrow{\sigma }\psi
\label{ppp6}
\end{equation}%
which depends only on the two components of the electron wave function in
carbon atoms.

In particular, using eq.(\ref{ss5}) for the first and second components of
the total wave function, we obtain for the current probability density
components $x$ and $y$%
\begin{equation}
j_{x}=\psi ^{\dag }\sigma _{x}\psi =v_{f}\sum\limits_{n=0}^{+\infty
}\sum\limits_{m=0}^{+\infty }\frac{c_{m}^{\ast }(t)c_{n}(t)e^{-i\omega
_{(n,m)}t}(-1)^{i}\alpha _{m}^{(i,j)}\alpha _{n}^{(i,j)}}{4L_{x}\sqrt{%
(\alpha _{n}^{2}+1)(\alpha _{m}^{2}+1)}}\left( \phi _{m}^{\ast }(\overline{%
\overline{y}})\phi _{n-1}(\overline{\overline{y}})+\phi _{m-1}^{\ast }(%
\overline{\overline{y}})\phi _{n}(\overline{\overline{y}})\right)
\label{ppp7}
\end{equation}%
and%
\begin{equation}
j_{y}=\psi ^{\dag }\sigma _{y}\psi =v_{f}\sum\limits_{n=0}^{+\infty
}\sum\limits_{m=0}^{+\infty }\frac{ic_{m}^{\ast }(t)c_{n}(t)e^{-i\omega
_{(n,m)}t}(-1)^{i}\alpha _{m}^{(i,j)}\alpha _{n}^{(i,j)}}{4L_{x}\sqrt{%
(\alpha _{n}^{2}+1)(\alpha _{m}^{2}+1)}}\left( \phi _{m-1}^{\ast }(\overline{%
\overline{y}})\phi _{n}(\overline{\overline{y}})-\phi _{m}^{\ast }(\overline{%
\overline{y}})\phi _{n-1}(\overline{\overline{y}})\right)  \label{ppp8}
\end{equation}%
Integrating the coordinates and taking the real part we finally obtain the
current probability in the $x$ and $y$ direction%
\begin{equation}
J_{x}(t)=\frac{v_{f}\alpha _{0}^{2}\alpha _{1}^{2}\beta \eta _{0}t}{2\hbar
(\alpha _{1}^{2}+1)(\alpha _{0}^{2}+1)}\left( 1-\lambda ^{2}\frac{\alpha
_{0}^{2}\alpha _{1}^{2}\beta ^{2}}{2\hbar ^{2}(\alpha _{0}^{2}+1)(\alpha
_{1}^{2}+1)}\eta _{0}^{2}t^{2}\right) \sin (\omega _{(1,0)}t)  \label{ppp9}
\end{equation}%
and%
\begin{equation}
J_{y}(t)=\lambda \frac{v_{f}\alpha _{0}^{2}\alpha _{1}^{2}\beta \eta _{0}t}{%
2\hbar (\alpha _{1}^{2}+1)(\alpha _{0}^{2}+1)}(1-\lambda ^{2}\frac{\alpha
_{0}^{2}\alpha _{1}^{2}\beta ^{2}}{2\hbar ^{2}(\alpha _{0}^{2}+1)(\alpha
_{1}^{2}+1)}\eta _{0}^{2}t^{2})\cos (\omega _{(1,0)}t)  \label{ppp10}
\end{equation}%
These results will be used in Section 5.

\end{document}